\documentclass{moriond}

\def\Journal#1#2#3#4{{#1} {\bf #2}, #3 (#4)}

\def\NPB{{\em Nucl. Phys.} B}

\def\PRL{\em Phys. Rev. Lett.}
\def\PRD{{\em Phys. Rev.} D}

\def\be{\begin{equation}}
\def\ee{\end{equation}}
\def\bea{\begin{eqnarray}}
\def\eea{\end{eqnarray}}

\newcommand{\Photo}{}

\usepackage{braket}
\usepackage{tensor}
\usepackage{amsmath}

\begin{document}
\vspace*{4cm}
\title{THE IMMIRZI FIELD IN MODIFIED GRAVITY: SUPERENTROPIC BLACK HOLES WITH SCALAR HAIR}

\author{S. BOUDET\footnote{On behalf of F. Bombacigno, G. Montani and M. Rinaldi. See\cite{Boudet2021} for reference.}}

\address{Dipartimento di Fisica, Università di Trento, Via Sommarive 14, I-38123 Povo (TN), Italy\\
and\\
Trento Institute for Fundamental Physics and Applications (TIFPA)-INFN,
Via Sommarive 14, I-38123 Povo (TN), Italy}

\maketitle\abstracts{In the context of $f(R)$ generalizations of the Holst action, endowed with a dynamical Immirzi field, we
derive an analytic solution describing asymptotically anti–de Sitter black holes with hyperbolic horizon.
These exhibit a scalar hair of the second kind, which ultimately depends on the Immirzi field radial
behaviour. In particular, we show how the Immirzi field modifies the usual entropy law associated to the
black hole. We also verify that the Immirzi field boils down to a constant value in the asymptotic region,
thus restoring the standard loop quantum gravity picture. We finally prove the violation of the reverse
isoperimetric inequality, resulting in the superentropic nature of the black hole, and we discuss in detail the
thermodynamic stability of the solution.}

\section{Introduction}
Loop Quantum Gravity (LQG) is one of the most prominent attempts towards the quantization of the gravitational interaction \cite{LQGRovelli,LQGThiemann}. It relies on the use of a specific set of variables, the Ashtekar-Barbero-Immirzi variables, in terms of which General Relativity (GR) can be formulated in a way more suitable for the application of known quantization techniques. One way of implementing the Ashtekar-Barbero-Immirzi variables in GR consists in modifying the Hilbert-Palatini (first order) action in the following way:
\begin{equation}
S[g_{\mu\nu},\Gamma\indices{^\mu_{\nu\rho}}] = \frac{1}{16\pi}\int \sqrt{-g} \left( \mathcal{R} - \frac{\gamma_0}{2} \varepsilon^{\mu\nu\rho\sigma}\mathcal{R}_{\mu\nu\rho\sigma} \right).
\end{equation}
Here $\mathcal{R}$ is the Ricci scalar defined in terms of a connection which is considered as an independent dynamical variable. The modification with respect to GR consists in the inclusion of the Holst term which does not affect the theory at the classical level by virtue of its on-shell vanishing due to the Bianchi identities. The coupling constant $\gamma_0$ is called the Immirzi parameter and it is known to be related to a quantum ambiguity \cite{ImmirziAmbiguity}. In particular, its value is arbitrary and does not appear in the classical theory, nevertheless it eventually ends up in the definition of quantum observables, e.g. the spectrum of the area operator:
\begin{equation}
\hat{A} \ket{\Psi} =\frac{1}{\gamma_0} \left( 8 \pi \ell_P^2 \sum_j \sqrt{j(j+1)} \right) \ket{\Psi}.
\end{equation}
In this way, it labels different quantum sectors of the theory which have been shown not to be related by a unitary transformation, constituting a purely quantum ambiguity which is still lacking a satisfactory and consistent resolution.
Some attempts in this direction have been made in \cite{Wang2017,NPBbbm21}, where the role of the Immirzi parameter in the loop quantization of scalar tensor theories has been investigated, while the first proposal actually consisted in determining $\gamma_0$ comparing the LQG computation of black hole entropy with the semiclassical Bekenstein-Hawking formula. Afterwards, some drawbacks of this idea were pointed out in \cite{Ghosh2011} and different numerical values for $\gamma_0$ were proposed in \cite{Pigozzo2021} with other motivations, thus leaving the question open.
In \cite{Taveras2008} a different idea was put forward. The authors proposed to promote the Immirzi parameter to a new fundamental field which may acquire a non trivial configuration in high energy regimes, e.g. in the primordial universe, and possibly relax to a non zero vacuum expectation value $\gamma_0$ during the cosmological evolution. Beside the original motivation, gravitaitonal models featuring the Immirzi field have been studied in different contexts yielding several interesting results (see\cite{Boudet2021} for a comprehensive list). In some of these works, the Immirzi field is considered within $f(R)$-like extensions of GR, where a general function of the Holst action is considered, possibly including a potential term for the Immirzi field as well:
\begin{equation}\label{action}
S[g_{\mu\nu},\Gamma\indices{^\mu_{\nu\rho}},\gamma] = \frac{1}{16\pi}\int \sqrt{-g}\left[ f\left( \mathcal{R} - \frac{\gamma(x)}{2} \varepsilon^{\mu\nu\rho\sigma}\mathcal{R}_{\mu\nu\rho\sigma} \right) - W(\gamma)\right].
\end{equation}
This is the gravitational model we will consider in the following.

\section{New exact black hole solution with scalar hair}
Despite several investigations on the Immirzi field already present in literature, there have been none regarding its effects in black hole spacetimes, this being the main subject of this note. The first difficulty one has to face in this direction comes from the existence of no-hair theorems, which make the search for vacuum spherically symmetric solutions different from GR ones extremely challenging. In a model like \eqref{action}, the consequence of this theorems is that a black hole solution must necessarily be characterized by the trivial constant configuration for the Immirzi field: $\gamma(x)\equiv \gamma_0$. Since this would amount to consider the usual case of a constant Immirzi parameter we are forced to bypass these theorems in order to highlight effects due to the dynamical nature of the Immirzi field. One way to overcome this difficulty consists in violating one of the hypothesis on which no-hair theorems stand. Here we chose to deal with the asymptotic flatness hypothesis, selecting the following Starobinsky-like function for the action
\begin{equation}
f(\chi)\sim \chi + \alpha \chi^2 -2\Lambda,
\end{equation}
where $\alpha$ is a free parameter and the inclusion of a cosmological constant $\Lambda$ allows for the existence of asymptotically (Anti)-de Sitter ((A)dS) spacetimes. The solution we are about to present is supported by the following choice for the Immirzi field potential:
\begin{equation}
W(\psi)=\frac{4\Lambda}{\text{csch}^2\left(\frac{\psi-\psi_0}{\sqrt{12}}\right) - 16\alpha \Lambda},
\end{equation}
where $\psi_0$ is another parameter of the theory and we introduced the field redefinition $\psi=\sqrt{3}\text{sinh}^{-1}\gamma$.
We report here a new, exact, asymptotically AdS solution describing a topological black hole with scalar hair. The metric is given by
\begin{equation}
ds^2 = \Omega(r) \left[ -h(r) dt^2 + h^{-1}(r) dr^2 + r^2 d\sigma^2 \right],
\end{equation}
where $d\sigma^2$ is the line element of a surface of negative constant curvature with area $\sigma$, while the two metric functions are defined by
\begin{equation}
h(r) = -\left(1+\frac{m}{r}\right)^2 + \frac{r^2}{l^2}, \qquad\qquad\qquad \Omega(r) = \frac{r(r+2m) + 48\alpha m^2/l^2}{(r+m)^2},
\end{equation}
with $l^2=-3/\Lambda$. The spacetime represents a black hole with mass $M=\sigma m/4\pi$ and event horizon located at
\begin{equation}
r_e = \frac{l}{2}\left( 1+ \sqrt{1+\frac{4m}{l}} \right).
\end{equation}
The scalar hair are of the secondary kind and are provided by the Immirzi field, which acquires the following non trivial radial behaviour:
\begin{equation}
\gamma(r) = \frac{e^{\psi_0/\sqrt{3}}(r+2m)^2 - e^{-\psi_0/\sqrt{3}} r^2 }{2r(r+2m)}.
\end{equation}
It relaxes at infinity to the constant value $\gamma_0\equiv \sinh{\psi_0/\sqrt{3}}$. Thus the standard picture with a constant Immirzi parameter is recovered asymptotically, while the presence of the black hole is able to excite a non trivial behaviour for the Immirzi field, in agreement with its original interpretation.

\section{Black hole thermodynamics}
\begin{figure}
\centerline{\includegraphics[width=0.6\linewidth]{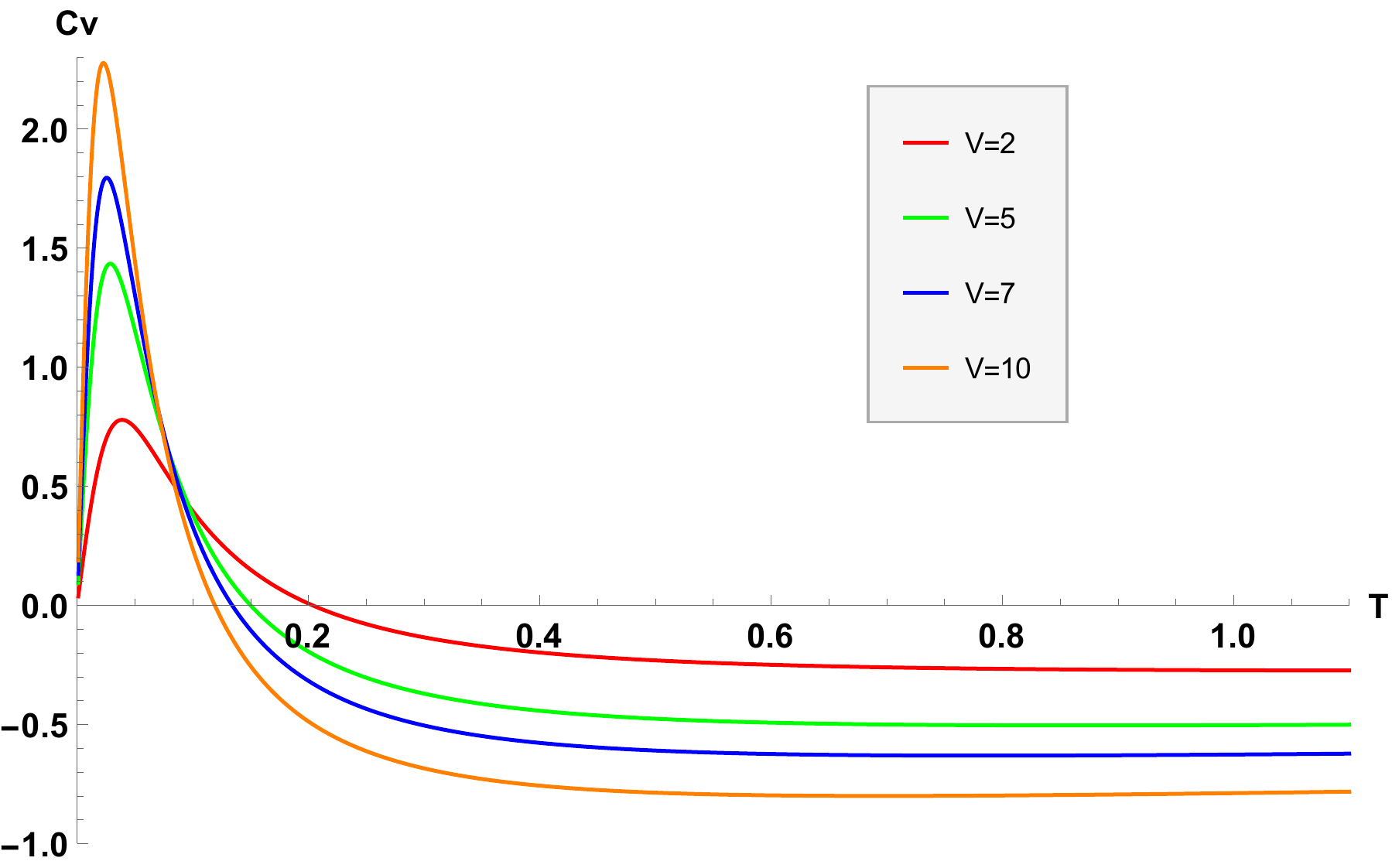}}
\caption[]{Specific heat at constant volume}
\label{fig:Cv}
\end{figure}

Although the non-flat asymptotics and the nontrivial horizon topology make this solution less prone to a direct astrophysical connotation, topological AdS black holes are of interest in the AdS/CFT context, where such solutions play a fundamental role \cite{Zaffaroni2020}. In this regard, it is important to investigate the thermodynamic properties of the black hole. Black hole thermodynamics has gained increasing attention since the seminal papers by Bekenstein and Hawking and by now several methods exist allowing to compute thermodynamic quantities of interest. The results presented here are obtained via euclidean path integral methods\footnote{We also checked the consistency of our results with the entropy obtained using Wald's formula}. The expression for the black hole entropy turns out to be given by
\begin{equation}
S=\left[ 1+4\alpha W(\psi_e) \right]\frac{A}{4},
\end{equation}
where $A$ is the event horizon area. The entropy is modified with respect to the usual Bekenstein-Hawking entropy formula, $S=A/4$, and the correction depends on the value the Immirzi field acquires on the black hole event horizon $\psi_e$. It is easy to see that, had we started a priori from a model with a constant Immirzi parameter, the entropy would have satisfied the standard area law $S=A/4$. Thus, the black hole entropy is one of the physical quantities which depends on the nature of the Immirzi field or parameter, allowing in principle to discern between these two cases. Further thermodynamic properties can be determined working in the so called extended phase space approach. It consists in enlarging the thermodynamic phase space beyond temperature and entropy, including also an effective pressure provided by the cosmological constant $P=-\Lambda/$ and its conjugate quantity, a thermodynamic volume $V$. This approach took the analogy between black holes and ordinary thermodynamic systems to its limits, revealing a plethora of different behaviours in common with ordinary matter systems (e.g. phase transitions, triple points, Van der Waals gas like behaviours) to the extent that it is now called black hole chemistry. Beside the similarities, also new contrasting features were highlighted. Some of these regard the properties of the thermodynamic volume $V$, which does not necessarily coincide with the geometric volume enclosed by the event horizon, i.e. $4\pi r_e^3/3$. Its peculiar properties were investigated resulting in a series of subsequent conjectures. The first one stated the validity of the Reverse Isoperimetric Inequality (RII), i.e.
\begin{equation}
\mathcal{R} \equiv \left(\frac{3V}{\sigma}\right)^{\frac{1}{3}}{\Big/} \left(\frac{A}{\sigma}\right)^{\frac{1}{2}}\geq 1.
\end{equation}
The inequality is saturated by the simplest AdS black hole, i.e. the Schwarzschild-AdS solution. This implies that the latter attains the maximum entropy possible for a given volume. The RII conjecture was motivated by the observation that all known solutions seemed to satisfy it but afterwards solutions were found violating the inequality. Since a solution with $\mathcal{R}<1$, has entropy greater than the one a Schwarzschild-AdS black hole would have, these black holes were dubbed superentropic black holes. Their properties in the AdS/CFT context were studied in \cite{Sinamuli2016,Xing2017}, opening interesting lines of investigation. A further conjecture has been proposed regarding the stability of superentropic black holes, stating that every superentropic solution must be thermodynamically unstable. Instabilities of this kind were analysed in literature computing the specific heats at constant pressure and volume. Regarding the solution presented here, it is easy to check that the RII is violated in almost all parameter space, resulting in the superentropic nature of the black hole. The specific heat at constant pressure turns out to be positive definite, while the specific heat at constant volume can become negative at high enough temperatures, as shown in Fig. \ref{fig:Cv}, thus signalling a thermodynamic instability, in line with the conjecture on the instability of superentropic black holes.
Sometime after the publication of \cite{Boudet2021}, a superentropic solution was found in \cite{Poshteh2021}, which seems to violate the conjecture, since both specific heats are positive. However, as a future perspective, it would be interesting to investigate again the thermodynamic stability of these solutions using more sophisticated, non-extensive methods \cite{Arcioni2005}, which have been shown to yield results in contrast with the conclusions drawn simply looking at the sings of the specific heats.

\section*{References}

\begin{thebibliography}{99}
\bibitem{LQGRovelli}C. Rovelli, {\em Quantum Gravity},
(Cambridge University Press, Cambridge, 2004).

\bibitem{LQGThiemann}T. Thiemann, {\em Canonical Quantum General Relativity},
(Cambridge University Press, Cambridge, 2007).

\bibitem{ImmirziAmbiguity}C. Rovelli and T. Thiemann, \Journal{\PRD}{57}{1009}{1998}.

\bibitem{NPBbbm21} F. Bombacigno {\it et al}, \Journal{\NPB}{963}{115281}{2021}.

\bibitem{Wang2017} O. J. Veraguth and Charles H. T. Wang, \Journal{\PRD}{96}{084011}{2017}.

\bibitem{Ghosh2011} A. Ghosh and A. Perez, \Journal{\PRL}{107}{241301}{ 2011}.

\bibitem{Pigozzo2021} C. Pigozzo {\it et al}, {\it Class. Quant. Grav.} {\bf 38}, 045001 (2021).

\bibitem{Taveras2008} V. Taveras and N. Yunes, \Journal{\PRD}{78}{064070}{2008}.

\bibitem{Boudet2021}
S. Boudet {\it et al}, \Journal{\PRD}{103}{084034}{2021}.

\bibitem{Zaffaroni2020} A.
Zaffaroni, {\it Living Rev. Relativ.} {\bf 23}, 2 (2020). 

\bibitem{Sinamuli2016} M. Sinamuli and R.B. Mann, {\it J. High Energ. Phys.} {\bf 2016}, 148 (2016).

\bibitem{Xing2017} Xing-Hui Feng and H. Lü, \Journal{\PRD}{95}{066001}{2017}.

\bibitem{Poshteh2021} M. B. J. Poshteh and R. B. Mann, \Journal{\PRD}{103}{104024}{2021}.

\bibitem{Arcioni2005} G. Arcioni and E. Lozano-Tellechea, \Journal{\PRD}{72}{104021}{2005}.

\end{thebibliography}
\end{document}